\documentclass[12pt]{article}
\usepackage{amsmath,amsfonts}
\usepackage[nosort]{cite}
\usepackage{graphicx}
\usepackage{psfrag}

\makeatletter\@addtoreset{equation}{section}\makeatother

\newcommand{\bea}{\begin{eqnarray}}
\newcommand{\eea}{\end{eqnarray}}
\newcommand{\ket}[1]{{|#1 \rangle}}
\newcommand{\vac}{|0\rangle}
\newcommand{\CN}{{\cal N}=4}
\newcommand{\tr}{\hbox{Tr}}
\renewcommand{\tilde}{\widetilde}
\renewcommand{\hat}{\widehat}

\newcommand{\gfrak}{{\mathfrak g}}

\renewcommand{\title}[1]{\vbox{\center\LARGE{#1}}\vspace{3mm}}
\renewcommand{\author}[1]{\vbox{\center#1}\vspace{3mm}}
\newcommand{\address}[1]{\vbox{\center\em#1}}
\newcommand{\email}[1]{\vbox{\center\tt#1}\vspace{3mm}}

\begin{document}

\begin{titlepage}

\begin{center}
\rightline{\tt NSF-KITP-09-61}

\bigskip\bigskip

\title{S-duality and the giant magnon dispersion relation}

\author{David Berenstein$^a$
and Diego Trancanelli$^{a,b}$}

\address{
$^a$Department of
Physics and $^b$Kavli Institute for Theoretical Physics, 
\\ University of California at Santa Barbara, CA 93106, USA}

\medskip\medskip

\email{dberens, dtrancan@physics.ucsb.edu}

\end{center}

\medskip\medskip

\abstract{\noindent
We use S-duality and planarity to propose an argument for
the non-renormalization of the dispersion relation of giant magnon
solutions in type IIB string theory on $AdS_5 \times S^5$. We
compute the spectrum of giant magnons for $(p,q)$-strings from field
theory at strong coupling by using the central charge properties of
electrically and magnetically charged supersymmetric states in the
Coulomb branch of ${\cal N}=4$ super Yang-Mills. We argue that the
coupling dependence of the giant magnon dispersion relation
conjectured in the literature using integrability assumptions is in
fact the only functional dependence compatible with S-duality.
}

\vfill

\end{titlepage}


\section{Introduction}

One of the most immediate, dynamical tests of the AdS/CFT
correspondence is provided by the fact that both type IIB string
theory and $\CN$ super Yang-Mills are believed to be exactly
invariant under $SL(2,\mathbb{Z})$ S-duality transformations. On the
string theory side of the correspondence, the S-operation acts on
the dilaton field by flipping its sign and interchanges, for
example, fundamental strings and D1-branes \cite{Hull:1994ys}. In
the gauge theory \cite{Montonen:1977sn}, it  trades a description
with gauge group $G$ and complexified coupling constant $\tau$ for a
description with group ${}^L G$ and coupling constant $-1/n_\gfrak\tau$.\footnote{${}^L G$ represents the
Langlands dual group and $n_\gfrak=1,2$ or $3$ depending on the choice of $G$. In this note we are interested in the case
$G=U(N)$, which is self-dual and has $n_\gfrak=1$. } Albeit a rigorous,
mathematical proof of these statements is still missing, a
convincing body of evidence in their favor has been produced over
the years.

The point of this note will be to assume that S-duality holds as
an exact symmetry in the AdS/CFT correspondence and to investigate
what this implies. In particular, we will be interested in
understanding what role S-duality plays in the integrability of the
sigma-model for $AdS_5\times S^5$ \cite{BPR} and the corresponding
integrable spin chain model in the dual ${\cal N}=4$ super
Yang-Mills \cite{MZ}.

Our main concern is to show that at strong coupling in the field
theory one can describe in detail not only the exact energies of some
fundamental string states, but also the energies of similar $(p,q)$-string states. The states we consider, the so-called {\it giant magnons} \cite{Hofman:2006xt}, are solitonic string solutions of
the string sigma model. The general study of such solutions was undertaken in \cite{Frolov:2002av}. In the spin chain limit these are described by perturbative gauge theory and become magnon excitations around some ferromagnetic ground state. 

Together with planarity arguments, this analysis will show that the giant magnon dispersion relation,
which was originally computed under various approximations in
\cite{Santambrogio:2002sb}, should not receive any perturbative
corrections, even though they are in principle allowed by
integrability \cite{Beisert:2005tm}. Indeed, in other AdS/CFT
setups \cite{ABJM}, such renormalizations are required to interpolate between
the weak and strong coupling limits \cite{Gaiotto:2008cg}. From this
point of view, the case of ${\cal N}=4 $ super Yang-Mills and its
$AdS_5\times S^5$ dual is rather special and deserves further
attention.

We will employ the approach of \cite{Berenstein:2005aa}, where part
of the gravitational geometry can be obtained from the distribution
of eigenvalues of a certain diagonal matrix model reduction of $\CN$
super Yang-Mills on $\mathbb{R}\times S^3$. The eigenvalues localize
on a 5-sphere that is identified with the $S^5$ of the dual
geometry, where the giant magnons live. The energies of the giant
magnons can be reproduced exactly by studying the off-diagonal modes
of the original field theory \cite{Berenstein:2005jq} that become
heavy for dynamical reasons and can be self-consistently integrated
out in the ground state. We will show that the eigenvalues can be
connected not only by fields carrying fundamental charges, but they
can also be connected by $(p,q)$-dyonic excitations carrying both
electric and magnetic charge and whose semiclassical energy will
exactly reproduce the energies of the giant magnons for the
$(p,q)$-strings. This will give us a spectrum of string states that
is manifestly covariant under S-duality. We will  moreover show that
compatibility of planarity with S-duality forces the corresponding
dispersion relation to be non-renormalized between weak and strong
coupling.


\section{Dyonic off-diagonal excitations}

It has been shown in \cite{Berenstein:2005aa} (see also
\cite{Berenstein:2005jq,Berenstein:2007zf} and
\cite{Berenstein:2008me} for a review) that one can truncate $\CN$
super Yang-Mills living on $\mathbb{R}\times S^3$ to a matrix
quantum mechanics of six commuting matrices given by the $s$-waves
of the scalar fields $\phi^I$. The truncation is argued to be valid
for low energy computations because  other modes become 
 heavy dynamically in the most probable
configurations. The dominance of commuting matrices can be shown
 analytically and numerically in various models
\cite{Berenstein:2008eg}.

In the truncation, one can diagonalize these commuting matrices
simultaneously and the (bosonic) eigenvalues that one obtains turn
out to localize on a 5-sphere with radius $r_0=\sqrt{N/2}$, where
$N$ represents the rank of the gauge group or, equivalently, the
number of eigenvalues. This localization is the result of a
competition between an attractive quadratic potential given by the
conformal coupling of the scalars to the curvature of the $S^3$ (a
mass term) and a repulsive interaction originating from the measure
Jacobian produced by the diagonalization.  In this setup, the gauge
dynamics can be considered to be spontaneously broken from $U(N)$ down to $U(1)^N$ to
account for the charges of the various states.

It is possible to study the off-diagonal excitations of this matrix
model in a perturbative way, by treating them as free harmonic
oscillators whose mass depends on the diagonal modes
\cite{Berenstein:2005jq}. A computation shows that the frequency of
these modes scales as the distance between pairs of eigenvalues.
More precisely, the off-diagonal element $\delta \phi^I_{ij}$ (with
$i\neq j$) has a mass squared
\bea
m^2_{ij} = 1 + \frac{\lambda}{4\pi^2} |\hat x_i-\hat x_j|^2\,,
\label{mass}
\eea
where the constant term comes from the curvature of the $S^3$,
$\lambda\equiv g^2_{YM}N= g_{YM}^2 (2r_0^2)$ is the 't Hooft coupling, and $\hat
x_i\equiv \vec x_i/|\vec x_i|$ is the unit-normalized 6-vector
formed by the $i$-th eigenvalues of the scalars $\phi^I$. The
off-diagonal modes can then be interpreted as fundamental string
bits connecting different eigenvalues on the $S^5$. The off-diagonal
fields carry $U(1)$ gauge charges and the Gauss constraint forces
them to be assembled into closed polygons, thereby providing a
combinatorial picture of closed strings as being formed by gluing
several open string bits into a loop. Notice that at strong coupling
the mass of these modes is very large and they can be integrated out of the
low-energy effective theory. In principle, the integration procedure
might generate additional interactions between the eigenvalues that
are not included by just performing the truncation to diagonal
configurations, so in general one could expect that the radius of
the sphere can be renormalized by these interactions to a new value $r\neq r_0$. This
changes
$\lambda$ in the expression (\ref{mass}) above to a more general function of the
coupling constants $\lambda$ and $\tau\equiv \theta/2\pi + i
4\pi/g^2_{YM}$
\bea
m^2_{ij} = 1 + \frac{ h(\lambda, \tau)}{4\pi^2}  |\hat x_i-\hat
x_j|^2\,,
\label{massren}
\eea
where $h(\lambda, \tau)$ is to be determined. This renormalization
would account for the expected renormalization of the giant magnon
dispersion relation.

We want to include magnetic charges in this picture
and consider the off-diagonal modes as bits of $(p,q)$-strings, and
not just of fundamental strings. In the planar limit and at strong
coupling the off-diagonal modes are not very sensitive to the
compactness of the sphere $S^3$, as their Compton wave-length
$l_C\sim 1/\lambda$ is much shorter than the curvature radius of the
space, so that the 3-sphere can be essentially replaced by flat space at
the Compton wave-length scale of the charged particles. This is true
so long as they can be considered to be heavy relative to the size
of the sphere, which requires a strong coupling limit in the 't
Hooft coupling. In this limit,  notice that configurations of
constant commuting matrices on the sphere become configurations on
the moduli space of flat directions of the ${\cal N}=4$ field theory
on flat space. The masses of these electric objects are not only
calculable, but they are protected by supersymmetry, since the
fundamental fields transform in short representations of
supersymmetry on flat space. The ${\cal N}=2$ central charge that
these states carry has to be identified with their electric charge
\cite{WO}, and the orientation and size is determined by the
expectation values of the scalar fields. Duality (together with
holomorphy) permits us not only to calculate the central charges and
masses of the fundamental charges, but also of magnetic and dyonic
charges. This was instrumental in the solution of ${\cal N}=2$ super
Yang-Mills \cite{SW}. For the case of ${\cal N}=4$ super Yang-Mills,
this was calculated by Sen \cite{Sen}.

It is immediate to compute the mass $\tilde m_{ij}$ of these
$(p,q)$ charged objects. This is obtained from (\ref{massren}) by
taking into account  the expression for the mass of the
$(p,q)$-dyons
\bea
T_{(p,q)}=T_{(1,0)}|p-q\tau|\,,\qquad
T_{(1,0)}=\frac{\sqrt{\lambda}}{2\pi}\,.
\label{tension}
\eea
Here we have written the $(p,q)$-dyon mass in terms of $\lambda$ and
$\tau$. One finds that $\tilde m_{ij}^2$ reads
\bea
\tilde m^2_{ij} = 1 + \frac{h(\lambda, \tau)
|p-q\tau|^2}{4\pi^2}  |\hat x_i-\hat x_j|^2\,.
\label{eq:tildem}
\eea
As already mentioned, the constant factor arises from the curvature
coupling of the scalar fields to the background metric of the $S^3$,
which should be the same for all BPS protected scalar particles if we enforce
S-duality. In general this constant gets replaced by $(\ell+1)^2$, where
$\ell$ is the orbital angular momentum quantum number on the $S^3$ \cite{Berenstein:2007zf}.
This term reproduces the bound state dispersion relation for giant
magnons \cite{Dorey}. This is also the momentum squared
operator on the sphere, which is also part of the dispersion
relation in flat space because of Lorentz symmetry.

The  regime where these calculations are valid is equivalent to
a decoupling limit where the sphere $S^3$ becomes of infinite
radius and the off-diagonal modes become BPS protected states
effectively living  in flat space. This requires first taking
large $N$, at strong 't Hooft coupling and then taking $\tau$ to be finite.


\section{The giant magnon dispersion relation}

We have now all the ingredients to analyze the dispersion relation
of the giant magnon of type IIB strings on $AdS_5\times S^5$. We
consider the solution corresponding to a $(p,q)$-string. Such string
has the same classical sigma model as a fundamental string
(described by a Nambu-Goto action)
\bea 
S_{(p,q)}=\frac{\sqrt{\lambda}}{2\pi}|p-q\tau| \int d^2\sigma
\, \sqrt{-\det g_{\alpha\beta}}\,, 
\eea
modulo a different overall factor given by the different tension of
the two objects (\ref{tension}). Being defined by the same
sigma-model, both a fundamental string and a $(p,q)$-string will
admit the same classical giant magnon solution, with just a
different dependence on $\tau$. One can then immediately generalize
the result of \cite{Hofman:2006xt} and write down the strong
coupling dispersion relation for giant magnons with both electric
and magnetic charge
\bea
(E-J)_{(p,q)}=\frac{\sqrt{\lambda} |p-q\tau|}{\pi}\left|
\sin\frac{k}{2}\right| \,,
\eea
where $k$ represents the world-sheet momentum of the giant magnon.

This can be justified at large (infinite) $N$ and finite coupling
because long classical D-strings that do not self-intersect can not
break in flat space. Remember that when $N$ is large the
$AdS_5\times S^5$ is very large in string units and can be replaced
by flat space at scales much larger than the string scale.

Now we shall compute the dispersion relation of giant magnon
solutions corresponding to $(p,q)$-strings. To this end, we would
need to consider a 2-impurity state with large momentum, which is
given at weak coupling by some BMN-like operator \cite{BMN}
\bea
\ket{k, J} \sim \sum_{\ell=0}^J e^{ 2\pi i k \ell/ J} \tr( Z^\ell
[X,Z] Z^{J-\ell} [Y,Z])\,.
\eea
At strong coupling, the $Z$'s are described by diagonal modes, whereas
$X$ and $Y$ are described by off-diagonal modes \cite{Berenstein:2005jq}.
Here, we include the possibility that the off-diagonal degrees of
freedom might be described by magnetic charges as well and we write
\bea
\ket{k,J}\sim \sum_{\ell=0}^J e^{2\pi i k \ell/J}\sum_{i,j=1}^N
z^\ell_i \,(M^\dagger)^i_{\; j}\, z^{J-\ell-2}_j\, (\tilde
M^\dagger)^j_{\; i}\vac.
\label{2imp}
\eea
By analogy, this can be thought of as a magnetic trace operator.
The $z$'s represent the (collective) coordinates of eigenvalues, while
the $M^\dagger$ and $\tilde M^\dagger$ represent the Fock space raising operators for the
corresponding states with given electric and magnetic
charges. This is allowed so long as one can argue that the
electric/magnetic impurities are well separated from each other and
that the system does not back-react substantially in their presence.
We assume this is consistent at this stage. One can argue this is
allowed by noticing that the $z$ variables are also excitations of
the eigenvalue degrees of freedom, so the charged objects are in a
sea of photon superpartners that can keep them apart from each other.
One can ignore the bound state problem if there are sufficiently
many such photons. Also, the force between eigenvalues due to the
presence of the charged particles is of order one, while the force
due to the collective repulsion of eigenvalues scales like a power
of $N$.
In the expression (\ref{2imp}) the sum over eigenvalue pairs ensures
that the discrete gauge symmetry of permutation of eigenvalues is
implemented. The vacuum also contains the information of the wave
function of the collective coordinate degrees of freedom.

Following \cite{Berenstein:2005jq}, the sum is done over $z$'s on the
sphere $S^5$ and it is interpreted as an element of the Fock space
of the off-diagonal modes. The sum has a very sharp maximum norm in
the amplitude for a fixed angle between the eigenvalues lying on a
diameter of the $S^5$. We can compute the energy of the state
(\ref{2imp}) and obtain the functional form of the dispersion
relation valid for arbitrary coupling
\bea
(E-J)_{(p,q)}=\sqrt{1+\frac{h(y,\tilde\tau)}{\pi^2}
\sin^2\frac{k}{2} }\,.
\label{dispersion}
\eea
Here $h(y,\tilde\tau)$, with $y\equiv 1/\lambda$ and
$\tilde\tau\equiv |p-q\tau|$, is an unknown function that we wish to
determine. We are using the variable $y$ rather than $\lambda$ to
stress the fact that we are expanding around $\lambda\to\infty$. An
important point is that, although we don't know {\it a priori} the
function $h(y,\tilde\tau)$, the emergent geometry approach we are
using guarantees that we will have the square root behavior for any
value of the coupling constant, as explained in
\cite{Berenstein:2005jq}.

The scope of this note is to demonstrate that the only functional
dependence compatible with the S-duality of ${\cal N}=4$ super
Yang-Mills is
\bea
h(y,\tilde\tau)=\frac{1}{y}  |p-q\tau|^2\,.
\label{claim}
\eea
This means that S-duality protects this
quantity from being renormalized, so that the (generalization to
$(p,q)$-strings of the) one-loop result of
\cite{Beisert:2005tm,Santambrogio:2002sb} is in fact exact to all
orders in perturbation theory.

The first step toward proving (\ref{claim}) makes use of the
emergent geometry approach described in the previous section. As
explained in \cite{Berenstein:2005jq}, the dependence on the
coupling constant appearing under the square root of the giant
magnon dispersion relation (\ref{dispersion}) is fixed by the
geometry of the eigenvalue distribution. In particular, it depends
on the mass of the off-diagonal modes and on the radius of the
5-sphere. In the case of the $(p,q)$-strings, the mass of the
off-diagonal modes (\ref{eq:tildem}) fixes
\bea
h(y,\tilde\tau)=f(y,\tau)  |p-q\tau|^2\,,
\label{h}
\eea
with $f(y,\tau)$ unknown.

At this point we use S-duality. We apply a transformation
\bea
\tau\to-\frac{1}{\tau}\,,\qquad y\to \frac{y}{|\tau|^2}\,,
\eea
to $h(y,\tilde\tau)$ in (\ref{h}) (which is the function
corresponding to a $(p,q)$-string) and set the result equal to the
$h(y,\tilde\tau)$ for a $(q,-p)$-string. S-duality maps in fact $
p\to q$ and $ q\to -p$. Redefining for convenience $g(u,v)\equiv u\,
f(u,v)$ we find that $g(y,\tau)$ has to satisfy
\bea
g\left(\frac{y}{|\tau|^2},-\frac{1}{\tau}\right)=g(y,\tau)\,.
\label{eq-g}
\eea
This is a modular equation whose only solution in the limit
$N\to\infty$ and $\lambda\to \infty$, we claim, is a constant. To
prove this we first need to invoke planarity. This is justified from
weak coupling to strong coupling by arguing that integrability
interpolates from the weak coupling spin chain to strong coupling by
summing planar diagrams only. Beyond that, there can be $1/N$
corrections, but these are ignored at large $N$.

In this limit the function $g(y,\tau)$ cannot depend on $\tau$
separately, so that (\ref{eq-g}) becomes
\bea
g\left(\frac{y}{|\tau|^2}\right)=g(y)\,.
\eea
Taking derivatives with respect to $y$ of this equation around
$y=0$, we see that all these derivatives have to be zero and
therefore $g$ is a constant. All of this assumes that the $y\to 0$
limit is smooth ({\it i.e.} $y=0$ is not an essential singularity),
as one would expect to be the case in the planar limit. Another way
to see this is by recalling that the $\theta$ angle cannot arise in
perturbation theory, as it is a non-perturbative effect given by
instantons. Our formula for the dispersion relation
(\ref{dispersion}) is inherently perturbative, thus we expect that
$g$ cannot depend on variations of $\theta$ around $y=0$ and then it
has to be constant.

This is valid in an expansion around $y=0$. We can extrapolate
this result to weak coupling, where we can compute the constant
value of $g$. This is equal to $g=1$
\cite{Beisert:2005tm,Santambrogio:2002sb}, thus proving
(\ref{claim}).


\section{Discussion}

We have provided an argument for the non-renormalization of the
dispersion relation of giant magnon solutions of type IIB strings
in $AdS_5 \times S^5$. Our proof is not based on diagramatic
techniques, but rather on bootstrapping S-duality, planarity, and a
certain matrix model arising in the strong coupling limit of $\CN$
super Yang-Mills on $\mathbb{R}\times S^3$. An important point to
stress is that this computation was possible because around
$N\to\infty$ and $\lambda\to \infty$ (the regime we have focused on
in this note) the regions of validity of the ``electric''
description and of its ``magnetic'' dual do in fact overlap, thus
making possible to apply S-duality. We have also identified the
complete spectrum of $(p,q)$-giant magnons in the process, obtaining
a collections of states that can be mapped into each other
consistently under S-duality.

Recently there has been much interest in the integrability of the
AdS$_4$/CFT$_{3}$ correspondence for M2-branes proposed in
\cite{ABJM} (see \cite{MZ2} and \cite{review} for some reviews). 
According to the logic followed here, 
one should not expect a similar non-renormalization
theorem to hold in that context. One has in fact no S-duality in
type IIA string theory (nor in M-theory) to protect the coupling
dependence of the dispersion relation of giant magnons in $AdS_4
\times \mathbb{C}P^3$, and one finds in fact that this quantity
depends on an interpolating function $h(\lambda)$, known only in the
weak and strong coupling limits
\cite{Gaiotto:2008cg,Berenstein:2008dc} and for which, very recently, an 
exact expression has been conjectured in \cite{Gromov:2014eha}. Similar considerations apply to the AdS$_3$/CFT$_2$ case; see \cite{Babichenko:2009dk}.


\subsection*{Acknowledgments}
This work was supported in part by the U.S. Department of Energy under grant DE-FG02-91ER40618 and NSF grants PHY05-51166 and PHY05-51164.



\begin{thebibliography}{99}
\addtolength{\parskip}{-1ex}

\bibitem{Hull:1994ys}
  C.~M.~Hull and P.~K.~Townsend,
  ``Unity of superstring dualities,''
  Nucl.\ Phys.\  B {\bf 438}, 109 (1995)
  [arXiv:hep-th/9410167];
  E.~Witten,
  ``String theory dynamics in various dimensions,''
  Nucl.\ Phys.\  B {\bf 443}, 85 (1995)
  [arXiv:hep-th/9503124].

\bibitem{Montonen:1977sn}
C.~Montonen and D.~I. Olive,
  ``Magnetic monopoles as gauge particles?,''
  Phys.\ Lett.\ B {\bf 72}, 177 (1977);
  P.~Goddard, J.~Nuyts and D.~I.~Olive,
  ``Gauge Theories And Magnetic Charge,''
  Nucl.\ Phys.\  B {\bf 125}, 1 (1977).

\bibitem{BPR}
  I.~Bena, J.~Polchinski and R.~Roiban,
  ``Hidden symmetries of the AdS(5) x S**5 superstring,''
  Phys.\ Rev.\  D {\bf 69}, 046002 (2004)
  [arXiv:hep-th/0305116].

\bibitem{MZ}
  J.~A.~Minahan and K.~Zarembo,
  ``The Bethe-ansatz for N = 4 super Yang-Mills,''
  JHEP {\bf 0303}, 013 (2003)
  [arXiv:hep-th/0212208];
  N.~Beisert and M.~Staudacher,
  ``The N=4 SYM Integrable Super Spin Chain,''
  Nucl.\ Phys.\  B {\bf 670}, 439 (2003)
  [arXiv:hep-th/0307042].

\bibitem{Hofman:2006xt}
  D.~M.~Hofman and J.~M.~Maldacena,
  ``Giant magnons,''
  J.\ Phys.\ A  {\bf 39}, 13095 (2006)
  [arXiv:hep-th/0604135].

\bibitem{Frolov:2002av}
  S.~Frolov and A.~A.~Tseytlin,
  ``Semiclassical quantization of rotating superstring in AdS(5) x S(5),''
  JHEP {\bf 0206}, 007 (2002)
  [arXiv:hep-th/0204226].

\bibitem{Santambrogio:2002sb}
  A.~Santambrogio and D.~Zanon,
  ``Exact anomalous dimensions of ${\cal N} = 4$ Yang-Mills operators with large R
  charge,''
  Phys.\ Lett.\  B {\bf 545}, 425 (2002)
  [arXiv:hep-th/0206079].

\bibitem{Beisert:2005tm}
  N.~Beisert,
  ``The $su(2|2)$ dynamic S-matrix,''
  Adv.\ Theor.\ Math.\ Phys.\  {\bf 12}, 945 (2008)
  [arXiv:hep-th/0511082].

\bibitem{ABJM}
  O.~Aharony, O.~Bergman, D.~L.~Jafferis and J.~Maldacena,
  ``N=6 superconformal Chern-Simons-matter theories, M2-branes and their
  gravity duals,''
  JHEP {\bf 0810}, 091 (2008)
  [arXiv:0806.1218 [hep-th]].

\bibitem{Gaiotto:2008cg}
  T.~Nishioka and T.~Takayanagi,
  ``On Type IIA Penrose Limit and N=6 Chern-Simons Theories,''
  JHEP {\bf 0808}, 001 (2008)
  [arXiv:0806.3391 [hep-th]];
  D.~Gaiotto, S.~Giombi and X.~Yin,
  ``Spin Chains in ${\cal N} = 6$ Superconformal Chern-Simons-Matter Theory,''
  arXiv:0806.4589 [hep-th];
  G.~Grignani, T.~Harmark and M.~Orselli,
  ``The SU(2) x SU(2) sector in the string dual of N=6 superconformal
  Chern-Simons theory,''
  Nucl.\ Phys.\  B {\bf 810}, 115 (2009)
  [arXiv:0806.4959 [hep-th]].

\bibitem{Berenstein:2005aa}
  D.~Berenstein,
  ``Large $N$ BPS states and emergent quantum gravity,''
  JHEP {\bf 0601}, 125 (2006)
  [arXiv:hep-th/0507203].

\bibitem{Berenstein:2005jq}
  D.~Berenstein, D.~H.~Correa and S.~E.~Vazquez,
  ``All loop BMN state energies from matrices,''
  JHEP {\bf 0602}, 048 (2006)
  [arXiv:hep-th/0509015].

\bibitem{Berenstein:2007zf}
  D.~Berenstein and S.~E.~Vazquez,
  ``Giant magnon bound states from strongly coupled ${\cal N} = 4$ SYM,''
  Phys.\ Rev.\  D {\bf 77}, 026005 (2008)
  [arXiv:0707.4669 [hep-th]].

\bibitem{Berenstein:2008me}
  D.~Berenstein,
  ``A strong coupling expansion for  ${\cal N} = 4$ SYM theory and other SCFT's,''
  Int.\ J.\ Mod.\ Phys.\  A {\bf 23}, 2143 (2008)
  [arXiv:0804.0383 [hep-th]].

\bibitem{Berenstein:2008eg}
  D.~E.~Berenstein, M.~Hanada and S.~A.~Hartnoll,
  ``Multi-matrix models and emergent geometry,''
  JHEP {\bf 0902}, 010 (2009)
  [arXiv:0805.4658 [hep-th]];
  T.~Azeyanagi, M.~Hanada, T.~Hirata and H.~Shimada,
  ``On the shape of a D-brane bound state and its topology change,''
  JHEP {\bf 0903}, 121 (2009)
  [arXiv:0901.4073 [hep-th]].

\bibitem{WO}
  E.~Witten and D.~I. Olive,
  ``Supersymmetry algebras that include
  topological charges,''
  Phys.\ Lett.\ B {\bf 78}, 97 (1978);
  H.~Osborn,
  ``Topological charges for $\CN$ supersymmetric gauge
  theories and monopoles of spin 1,''
  Phys.\ Lett.\ B {\bf 83}, 321 (1979).

\bibitem{SW}
  N.~Seiberg and E.~Witten,
  ``Monopole Condensation, And Confinement In N=2 Supersymmetric Yang-Mills
  Theory,''
  Nucl.\ Phys.\  B {\bf 426}, 19 (1994)
  [Erratum-ibid.\  B {\bf 430}, 485 (1994)]
  [arXiv:hep-th/9407087].

\bibitem{Sen}
  A.~Sen,
  ``Dyon - monopole bound states, selfdual harmonic forms on the multi -
  monopole moduli space, and $SL(2,\mathbb{Z})$ invariance in string theory,''
  Phys.\ Lett.\  B {\bf 329}, 217 (1994)
  [arXiv:hep-th/9402032].

\bibitem{Dorey}
  N.~Dorey,
  ``Magnon bound states and the AdS/CFT correspondence,''
  J.\ Phys.\ A  {\bf 39}, 13119 (2006)
  [arXiv:hep-th/0604175].

\bibitem{BMN}
  D.~E.~Berenstein, J.~M.~Maldacena and H.~S.~Nastase,
  ``Strings in flat space and pp waves from N = 4 super Yang Mills,''
  JHEP {\bf 0204}, 013 (2002)
  [arXiv:hep-th/0202021].

\bibitem{MZ2}
  J.~A.~Minahan and K.~Zarembo,
  ``The Bethe ansatz for superconformal Chern-Simons,''
  JHEP {\bf 0809}, 040 (2008)
  [arXiv:0806.3951 [hep-th]];
  N.~Gromov and P.~Vieira,
  ``The all loop AdS4/CFT3 Bethe ansatz,''
  JHEP {\bf 0901}, 016 (2009)
  [arXiv:0807.0777 [hep-th]].
  
  \bibitem{review}
T.~Klose,
  ``Review of AdS/CFT Integrability, Chapter IV.3: N=6 Chern-Simons and Strings on AdS4xCP3,''
  Lett.\ Math.\ Phys.\  {\bf 99}, 401 (2012)
  [arXiv:1012.3999 [hep-th]];
   A.~E.~Lipstein,
  ``Integrability of N = 6 Chern-Simons Theory,''
  arXiv:1105.3231 [hep-th].

\bibitem{Berenstein:2008dc}
  D.~Berenstein and D.~Trancanelli,
  ``Three-dimensional N=6 SCFT's and their membrane dynamics,''
  Phys.\ Rev.\  D {\bf 78}, 106009 (2008)
  [arXiv:0808.2503 [hep-th]].
  
  \bibitem{Gromov:2014eha} 
  N.~Gromov and G.~Sizov,
  ``Exact Slope and Interpolating Functions in ABJM Theory,''
  arXiv:1403.1894 [hep-th].
  
  \bibitem{Babichenko:2009dk} 
  A.~Babichenko, B.~Stefanski, Jr. and K.~Zarembo,
  ``Integrability and the AdS(3)/CFT(2) correspondence,''
  JHEP {\bf 1003}, 058 (2010)
  [arXiv:0912.1723 [hep-th]].

\end{thebibliography}
\end{document}